\documentclass[aps,twocolumn,superscriptaddress,footinbib,longbibliography,prl]{revtex4-1} 

\usepackage{longtable}
\usepackage{morefloats}
\usepackage[dvips]{graphicx}
\usepackage{color}
\usepackage{epsfig,graphicx,amsfonts,amsbsy}
\usepackage{amsmath,amsfonts,amsthm,amssymb}
\usepackage{appendix}
\usepackage{makeidx}
\usepackage{url}
\usepackage{verbatim}
\usepackage[bookmarksnumbered,pdfpagelabels=true,plainpages=false,colorlinks=true,linkcolor=blue,citecolor=red,urlcolor=blue]{hyperref}
\usepackage[rightcaption]{sidecap}
\usepackage{array}
\usepackage{multirow}
\usepackage{tabularx}
\usepackage{braket} 
\usepackage{dsfont}
\usepackage{bbold}
\usepackage{etoolbox}

\usepackage{dcolumn}
\usepackage{bm}
\usepackage{blindtext}
\usepackage{comment}
\usepackage{physics}

\def \fcfm {Departamento de F\'isica, FCFM, Universidad de Chile, Santiago, Chile.}
\def \uai {School of Engineering and Sciences, Universidad Adolfo Ibañez, Santiago, Chile}
\def \usach {Departamento de F\'isica, Universidad de Santiago de Chile, 9170124, Santiago, Chile.}
\def \cedenna {Centro  de Nanociencia y Nanotecnología CEDENNA, Avda. Ecuador 3493, Santiago, Chile.}

\begin{document}
\title{Magnonics along the wall in  Bimeron Chain Domain Walls}
\author{Carlos Saji}
\affiliation{\fcfm}
\author {Eduardo Saavedra}
\affiliation{\usach}
\author{Roberto E. Troncoso}
\affiliation{\uai}
\author{Mario A. Castro}
\affiliation{\fcfm}
\author{Sebastian Allende}
\affiliation{\usach}
\affiliation{\cedenna}
\author{Alvaro S. Nunez}
\affiliation{\fcfm}
\date{\today}


\begin{abstract}
We demonstrate that domain walls built from bimeron chains (bc-DW) in two-dimensional systems constitute a spontaneously assembled medium that holds magnonic excitations along its direction. We prove that such magnons are topological, leading to protected edge states. We also verify the stability of the domain walls and its edge modes' resilience against disorder. Analytical calculations and micromagnetic simulations support our findings. The robustness of these edge modes holds promise for potential applications in the design of nanoscale magnonic devices for information storage and transport.
\end{abstract}

\maketitle


{\it Introduction.}-- Spin waves, whose quanta are magnons, have attracted significant attention in recent years \cite{Flebus2023, Roldan2017, Doornenbal2019, HidalgoSacoto2020, Pirro2021, Harms2024}. These are collective spin excitations in magnetically ordered materials, such as ferromagnets, ferrimagnets, and antiferromagnets \cite{Kittel1987, Kittel2004}. Due to their low energy consumption and long coherence lengths, spin waves and magnons are considered promising candidates for next-generation information carriers.  

In a parallel development, topological band theory became a prominent area of research in condensed matter physics \cite{Haldane2017, Kosterlitz2017, Moessner2021, Hasan2010, Qi2011}.  In the past decade, research on topological matter has gained substantial momentum. Topological matter exhibits non-trivial structures characterized by gapped behavior in the bulk and robust conducting states on their surfaces or edges. These edge or surface states hold significant promise for practical applications. Beyond electronic states, topological behavior has also been observed in other types of excitations. For example, photons \cite{Price2022}, phonons \cite{Xu2024}, polaritons \cite{Karzig2015}, and plasmons \cite{Dai2020}, among others, have been shown to host topological properties. From this perspective, topological magnonic systems and the nascent field of topological magnonics are precipitating a new range of effects, with potential applications in energy-efficient information technology \cite{Shindou2013, RoldanMolina2016, Aguilera2020, Costa2020, Jaeschke2021, Wang2021, McClarty2022, dosSantosDias2023, Saji2023, Tapia2024}.
\begin{figure}[h!]
\centering
\includegraphics[width=0.7\columnwidth]{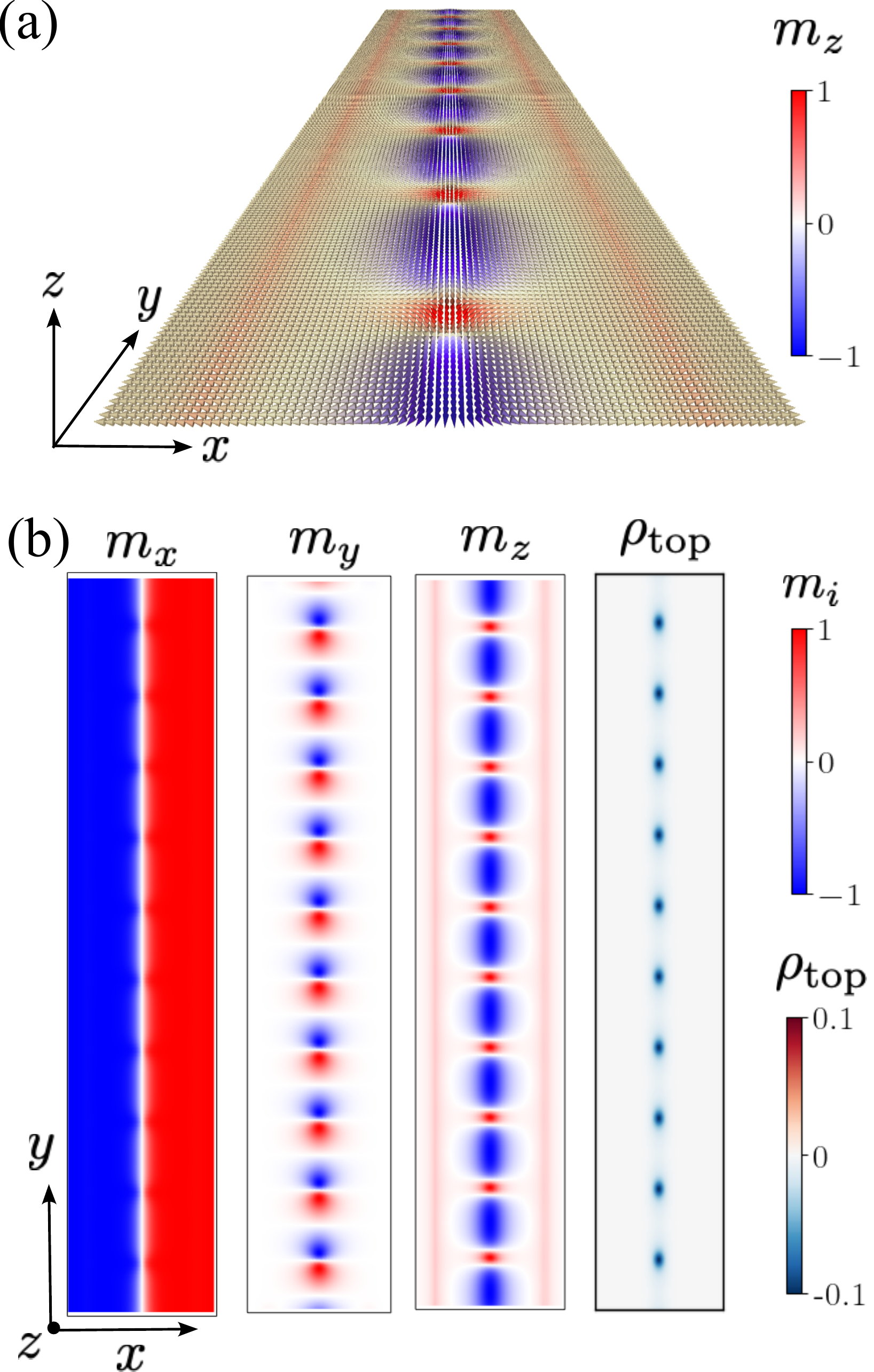}
\caption{{Visualization of the bimeron chain domain wall.} (a) Magnetization texture in the nanostripe, containing $10$ bimerons distributed along the DW, and obtained through  micromagnetic simulations in $\mathrm{Mumax}^{3}$.(b) Cartesian components of the magnetization and the topological density charge $\rho_{\mathrm{top}}= \bm{m}\cdot (\partial_{x}\bm{m} \times \partial_{y}\bm{m}) \ [\mathrm{nm}^{-2}]$, where each bimeron has a topological charge $\mathcal{Q}=-1$.}
\label{fig:Bimeron_Chain} 
\end{figure}
    
Bimerons are topological spin textures stabilized by in-plane anisotropy in non-centrosymmetric ferromagnetic materials, exhibiting several properties with skyrmions \cite{Kharkov2017, Gobel2019, Gao2019, Gobel2021, Castro2023}. In particular, they can form lattice-like structures and have recently been observed in open or closed chain formations \cite{Mukai2024, Zhang2020}.  Bc-DWs have been highlighted for their distinctive dynamical behaviors, such as serving as a track for bimeron motion and their ability to mitigate the skyrmion Hall effect \cite{Chen2024}. Additionally, domain-wall bimerons have been experimentally observed in various studies \cite{Nagase2021, Li2021, Amari2024}. Recent research has also explored the interaction between spin waves and bimerons \cite{Liang2023}. However, the topological phenomena associated with magnons in the vicinity of a bc-DW remain largely unexplored.

In this Letter we demonstrates, by performing micromagnetic simulations and analytical calculations of the magnonic states, that domain walls built based on bimeron chains in two-dimensional systems constitute a spontaneously assembled medium that, acting as a tunable magnonic metamaterial, holds topological edge modes. Although bulk modes function as magnonic waveguides, the robustness of these edge modes holds promise for potential applications in the design of nanoscale devices for information storage and transport. In addition, we observe that bc-DWs and their topological edge modes remain stable against substantial magnetic disorder,
introduced into the system in the form of Voronoi grains, highlighting their potential use in the design of spintronic devices. 
Our work bridges the gap in understanding topological phenomena associated with spin-wave modes of bc-DWs, providing valuable insights into the potential of these systems for hosting topologically protected magnonic modes.

{\it Theoretical model.}-- We consider a chiral thin ferromagnetic material with in-plane magnetic anisotropy along the $x$-axis, with a rectangular geometry of dimensions $L_{x}$ and $L_{y}$, as in Fig. \ref{fig:Bimeron_Chain}. The magnetic energy density reads, $\mathcal{E}= \mathcal{E}_{\mathrm{ex}} + \mathcal{E}_{\mathrm{DMI}}  + \mathcal{E}_{K} +  \mathcal{E}_{\mathrm{Z}} + \mathcal{E}_{\mathrm{DDI}}$, where $\mathcal{E}_{\mathrm{ex}}= A(\nabla \boldsymbol{m})^2$ represents the Heisenberg exchange, $\mathcal{E}_{\mathrm{DMI}}= D(m_z(\nabla \cdot \boldsymbol{m})-(\boldsymbol{m} \cdot \nabla) m_z)$ the interfacial Dzyaloshinskii–Moriya (DMI), $\mathcal{E}_{K}=K_{u}(1-m_{x}^2)$ the in-plane easy-axis anisotropy along the ${x}$-direction, and $\mathcal{E}_{\mathrm{Z}}=-B_{y}{m_y}$ the Zeeman coupling $B_{y}$ a static global in-plane field. The dipolar field energy, $\mathcal{E}_{\mathrm{DDI}}=-\mu_0 M_{s} \boldsymbol{H}_{\boldsymbol{d}} \cdot \boldsymbol{m}/2$, is approximated as an effective anisotropy and a confining harmonic well for the domain wall. To pin the domain wall and prevent rigid displacements of the chain, we introduce an easy-axis surface anisotropy $K_{S}$ along the $x$-direction at the boundaries $x=\pm L_{x}/2$. Let us introduce the polar representation (oriented to the $x$-axis), 
$\boldsymbol{m} = \left(\cos \Theta, \sin \Theta \sin \Phi, \sin\Theta \cos \Phi\right)$, and we consider the following ansatz for the bc-DW,
\begin{align}\label{eq:ansatz}
\Theta(x,y)= 2 \arctan\left[e^{-x/w +X(y)}\right], \ \   \Phi(x,y)= \eta(y),
\end{align}
which corresponds to a set of domain walls (parametrized by the continuous $y$ variable) with core position $w X(y)$ and chiral angle $\eta(y)$. The thickness of the DW is determined by $w=\sqrt{A/(2K_{\mathrm{eff}})}$ \cite{Kim2023}, where the effective anisotropy $K_{\mathrm{eff}}= K_u+2\mu_{0}M_{s}^{2}$ takes into account a significant contribution from the dipolar field.  

Next, we determine the domain wall profile using $X(y,t)$ and $\eta(y,t)$ as collective variables, and by substituting the ansatz Eq. (\ref{eq:ansatz}) into the magnetic energy density, we arrive at the effective energy of the bc-DW,
\begin{align}
E[X,\eta]&\nonumber= \int  \left[ w A\left ( (\partial_{y}X)^{2} + (\partial_{y}\eta)^{2} \right )\right.  \\
&\left.  \qquad + D\cos(\eta) + w B_{y}\sin(\eta) + U(X)  \right] \ dy \label{eq:effective_energy},    
\end{align}
where $U(X)$ represents a pinning potential considering the shape anisotropy and the dipolar field on the bc-DW. Now, considering small deviations from the equilibrium position $X=0$, we can model $U(X)$ as a one-dimensional harmonic potential $U(X)= k_{el}X^{2}$. To describe the effective dynamics, we determine the Lagrangian of the collective variables, $\mathcal{L}[X,\eta]={\cal K}[X,\eta] - \gamma_{0} E[X,\eta] $, with the kinetic energy, ${\cal K}[X,\eta] =\int G_{X\eta}\ X\partial_{t}\eta \ dy$, written in terms of the gyrotropic tensor element between $X$ and $\eta$, 
\begin{align} 
G_{X\eta}=&\nonumber  \int  \sin\Theta \left( \partial_{X}\Theta \partial_{\eta}\Phi - \partial_{\eta}\Theta \partial_{X}\Phi \right) \ d^{2}x \\
=& w \int  \sin\Theta \partial_{x}\Theta \ dx  =  2 w,  \label{eq:gyrotropic}
\end{align}
Thus, we obtain the Lagrangian for the collective dynamical conjugate variables,
 \begin{align}
\mathcal{L}&[X,\eta]
=\nonumber \int [ 2 w  X\partial_{t}\eta  - \gamma_{0} w A\left ( (\partial_{y}X)^{2} + (\partial_{y}\eta)^{2} \right )\\
&\quad   - \gamma_{0} D\cos\eta - \gamma_{0} w B_{y}\sin\eta - \gamma_{0}k_{el}X^{2} ] \ dy.
\end{align}
The Euler-Lagrange equation for $X$ leads to $w \partial_{t}\eta= -\gamma_{0} w A\partial_{y}^{2}X + \gamma_{0} k_{el}X$, which we solve using the Green function $\mathcal{G}= \left(  k_{el} - w A\partial_{y}^{2} \right)^{-1} $ as $X=\gamma_{0}^{-1}w\ \mathcal{G}\ \partial_{t}\eta$. Under the assumption
that the stiffness $k_{el}$ is strong enough such that $w A G_{y}^{2} \ll k_{el}$, where $G_{y}= 2\pi/\lambda$ and $\lambda$ represent the period of the bc-DW; thus the Green’s function becomes sufficiently local in space, and we can make the
approximation $X= w/(\gamma_{0} k_{el}) \ \partial_{t}\eta$. Substituting it into $\mathcal{L}[X,\eta]$, the new effective Lagrangian density is given by $\mathcal{L}_{\mathrm{eff}}[\eta]=  m_{\mathrm{eff}}(\partial_{t}\eta)^{2}- (\partial_{y}\eta)^{2} - 2 d \cos\eta
- 2 b\sin\eta$, with the effective mass $m_{\mathrm{eff}}= w/(\gamma_{0}^{2} k_{el}A)$, $d= D/(2wA)$, and $b= B_{y}/(2A)$. \begin{figure}[h!]
	\centering
	\includegraphics[width=\columnwidth]{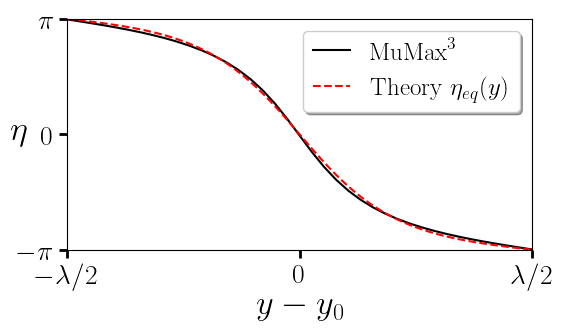}
	\caption{{Chiral angle $\eta$ of a bc-DW along the DW symmetry axis $x=0$} of Fig. \ref{fig:Bimeron_Chain}. In the solid line is depicted the value of $\eta$ according to the micromagnetic simulations in $\mathrm{Mumax}^{3}$, whereas in the dashed red line is plotted the theoretical prediction given at Eq. (\ref{eq:Eta_theorical}), with $\kappa=0.99$ and $b=0$. Analytical results are well consistent with the micromagnetic simulations.}
	\label{fig:Eta_plot}
\end{figure}
Therefore, the equilibrium configuration satisfies the so-called sine-Gordon differential equation,
\begin{equation}\label{eq:sineGordon}
	\frac{d^{2} \eta_{eq} }{d y^{2}}= - \epsilon\sin(\eta_{eq}-\delta_{b}), 
\end{equation}	
with $\epsilon= \sqrt{d^{2}+b^{2}}$ and $\delta_{b}= \arctan(b/d)$. The solution of the differential equation in Eq. (\ref{eq:sineGordon}) can be expressed in terms of the Jacobi amplitude function, $\mathrm{am}$ \cite{Akhiezer1990, Whittaker1996},
\begin{equation}\label{eq:Eta_theorical}
	\eta_{eq}(y) = -2\  \mathrm{am}\left( \sqrt{\frac{\epsilon}{\kappa}} (y-y_{0}) ,\kappa \right)  + \delta_{b},
\end{equation}
where  $\kappa$ stands for the elliptic modulus. The solution at Eq. (\ref{eq:Eta_theorical}) defines a periodic magnetization profile $m_{z}(0,y)=\cos(\eta_{eq}(y))$ and $m_{y}(0,y)=\sin(\eta_{eq}(y))$, with the period given by $\lambda=2 K(\kappa)  \sqrt{\kappa/\epsilon}$, where $K(\kappa)= \int_{0}^{\pi/2} (1-\kappa^{2}\sin^{2}\theta)^{-1/2} d\theta$ denotes the complete elliptic integral of the first kind. More detailed information between the relation of $\lambda$ and $\kappa$ are found at the Supplementary Material. On the other hand, the constant $y_{0}$ is determined by the Neumann boundary conditions \cite{Rohart2013},
\begin{equation}\label{eq:boundary_condition}
    \frac{\partial \boldsymbol{m}}{\partial \boldsymbol{n}}=\frac{D}{2 A}(\hat{\boldsymbol{z}} \times \boldsymbol{n}) \times \boldsymbol{m},
\end{equation}
where $\boldsymbol{n}$ denotes the exterior normal vector to the boundary in the $XY$ plane. Applying Eq. (\ref{eq:boundary_condition}) with $\boldsymbol{n}=-\hat{\boldsymbol{y}}$ at $y=0$, we obtain $\eta'(0)= D/(2A)$, hence $\sqrt{{\epsilon}/{\kappa}}\,\mathrm{am}'\left( - \sqrt{{\epsilon}/{\kappa}}\, y_{0} ,\kappa \right)= D/(2A)$, which can be solved numerically.  
We compare this result with the one obtained using $\mathrm{Mumax}^{3}$ for the chiral angle, see Fig. \ref{fig:Eta_plot}.  

{\it Spin wave spectrum.}-- We now focus in the study of small perturbations around the equilibrium solution $\eta_{eq}$. Expanding the Lagrangian $\mathcal{L}_{\mathrm{eff}}[\eta]$ up to second order in $\delta \eta= \eta-\eta_{eq}$, we find the dynamics of linear excitations, which are determined by the eigenvalue problem,
\begin{equation}\label{eq:eigenvalues_problem}
 \mathcal{H} \Psi =  \left( -\partial_{y}^{2} + V_{\mathrm{eff}}(y) \right) \Psi =  m_{\mathrm{eff}} \omega^{2}\ \Psi,
\end{equation}
where $\Psi(\omega)$ is the Fourier transform of $\delta \eta(t)$. The effective potential $V_{\mathrm{eff}}(y)$ is defined by,
\begin{equation}\label{eq:effective_potential}
	V_{\mathrm{eff}}(y) = - \epsilon \cos \left[ 2\  \mathrm{am}\left(  \sqrt{\frac{\epsilon}{\kappa}} (y-y_{0}) ,\kappa \right) \right] .
\end{equation}
We observe that it is a periodic potential with a period of $\lambda$. Furthermore, if $B_y=0$, then we have that $V_{\mathrm{eff}}(y)=- \epsilon \cos(\eta_{eq}(y))= - \epsilon\ m_{z}(x=0,y)$ attains its minimum and maximum values precisely at $m_{z}=1$ and $m_{z}=-1$, respectively. This allows us to interpret the effective dynamics as analogous to those of a particle moving in a one-dimensional crystal with potential wells localized at the core of the bimerons of the chain. 
 
  \begin{figure}[h!]
 	\begin{center}
 		\includegraphics[width=0.7\columnwidth]{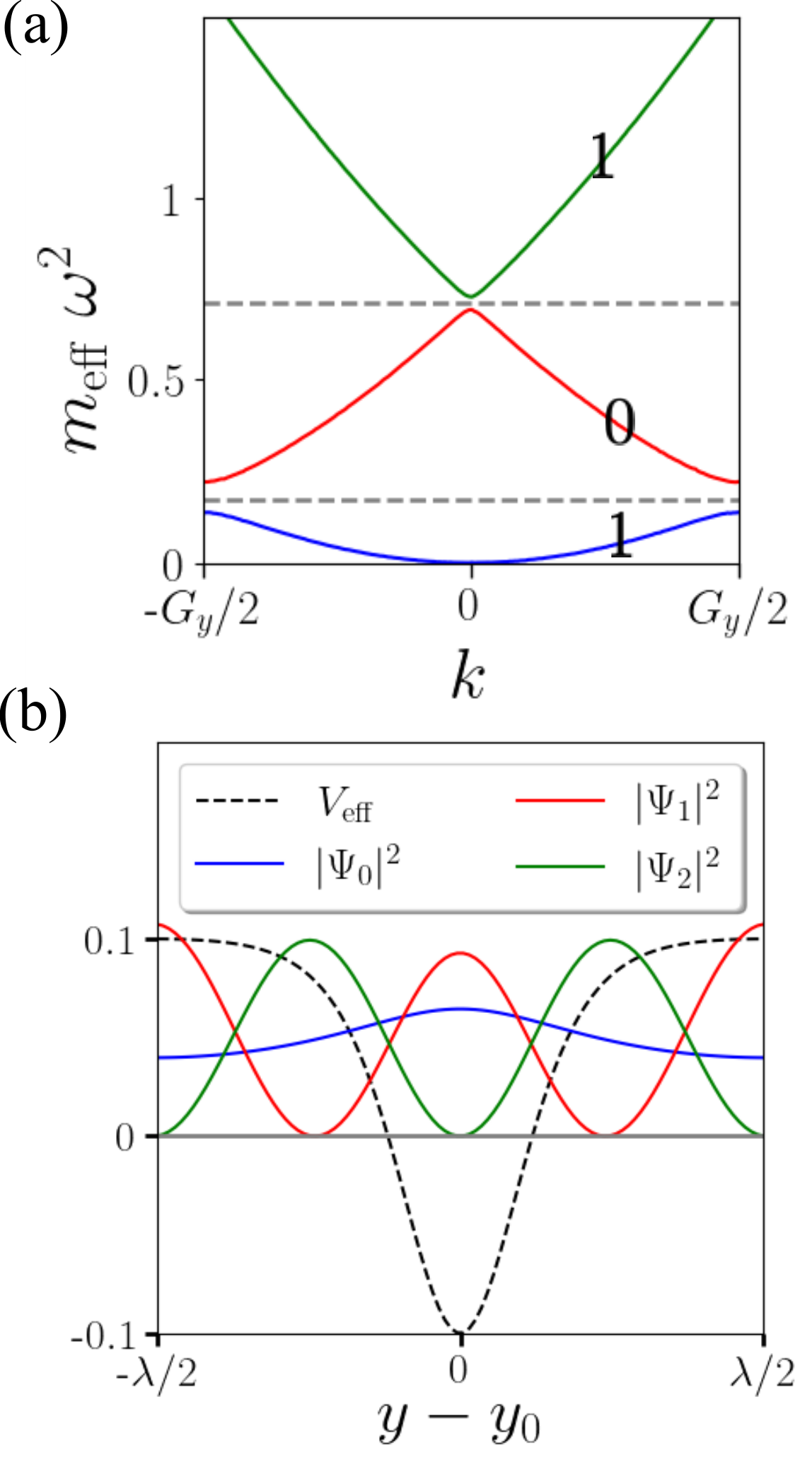}
 		\caption{
        (a) Spectrum of the Hamiltonian $\mathcal{H}(k)$ obtained by solving the eigenvalue problem in Eq. (\ref{eq:problem_momentum}). The effective potential $V_{\mathrm{eff}}$ is defined in Eq. (\ref{eq:effective_potential}) with parameters $\kappa=0.99$ and $\epsilon=0.1$. The respective Zak phase $\gamma_{n}\in \left\lbrace 0,1 \right\rbrace $ is marked on each band. The horizontal dashed lines represent topological band gaps. We note that the spectrum is indeed positive. (b) Normal modes $\Psi_{n}(y)$ calculated by solving Eq. (\ref{eq:problem_momentum}) with periodic boundary conditions ($k=0$). In color, solid lines show the amplitude of the mode $\Psi_{n=0,1,2}$. The dashed line represents the effective potential at Eq. (\ref{eq:effective_potential}) with parameters $\kappa=0.99$ and $\epsilon=0.1$.}
 		\label{fig:Bands}
 	\end{center}
 \end{figure}
 
Let us focus on the spectrum of the Hamiltonian $\mathcal{H}$. We first note that there is a zero energy mode $\Psi_{0}(y)$, which is described by $\Psi_{0}(y)= \eta_{eq}'(y)$. This mode is the Goldstone mode associated with the translational symmetry along the $y$-axis.  Separately, stabilizing the equilibrium state defined by $\eta_{eq}$ requires that all eigenvalues in Eq. (\ref{eq:eigenvalues_problem}) be non-negative, $\varepsilon_{n}=m_{\mathrm{eff}} \omega_{n}^{2}\geq 0$. In fact, following the supersymmetric quantum mechanics approach (SUSY-QM) \cite{Sukumar1985, Cooper1995}, the Hamiltonian (\ref{eq:eigenvalues_problem}) can be written as $\mathcal{H}=A^{\dagger}A$, where the operator $A$ is defined by $A=-\partial_{y}+W(y)$ with the superpotential $W(y)=\eta_{eq}''(y)/\eta_{eq}'(y)$. This implies that the $\mathcal{H}$ spectrum is nonnegative. See the Supplementary Material for more details. This technique has previously been used in magnonics \cite{Gonzalez2010, Lee2022}.   Moreover, using the SUSY-QM correspondence, we can calculate an approximate solution for the Bloch eigenfunctions of $\mathcal{H}(k)$ in the asymptotic case $\kappa \approx 1$, we obtain the following result,
\begin{align} 
	\Psi_{n,k}(y) &\approx \left( W(y) - i (k+n G_{y})  \right) e^{i (k+n G_{y}) (y-y_{0})}\\
	m_{\mathrm{eff}} \omega_{n,k}^{2} &\approx (k+n G_{y})^{2}+\epsilon.
\end{align}

In the general case, we determine the band structure through the exact diagonalization procedure as follows: we begin by writing the Fourier expansion of a Bloch wave $\Psi_{k}(y)= \frac{1}{\sqrt{N}} \sum_{n=-N}^{N} e^{i(k+n G_{y})(y-y_{0})} \psi_{n}$, where $N\to \infty$. Substituting in Eq. (\ref{eq:eigenvalues_problem}), the secular equation reads,
\begin{align}\label{eq:problem_momentum}
	\sum_{m} \left[ \left( k+ n G_{y} \right)^{2} \delta_{n m} + V_{n-m} \right] \psi_{m}= m_{\mathrm{eff}} \omega^{2} \ \psi_{n},
\end{align} 
where $V_{n}=\frac{1}{\lambda} \int_{-\lambda/2}^{\lambda/2} e^{i n G_{y} (y-y_{0})}V_{\mathrm{eff}}(y) \ dy$ are the Fourier coefficients of $V_{\mathrm{eff}}$, which we calculate in the Supplementary Material. The above system is solved numerically, from which we illustrate the lower-energy normal modes $\Psi_{n=0,1,2}(y)$ with periodic boundary conditions ($k=0$), and their associated band spectrum in Fig. \ref{fig:Bands}.  

The topology of the bands is characterized by the Zak phase \cite{Zak1989}, defined by the integral of the Berry connection \cite{Berry1984} over the Brillouin zone (BZ),
$$
\gamma_{n}= \frac{i}{\pi}\oint_{BZ}  \bra{ \Psi_{n}(k)}\ket{\frac{\partial }{\partial k } \Psi_{n} (k)}  \ dk \ (\mathrm{mod}\ 2)
$$
The matrix Hamiltonian $\mathcal{H}(k)$ defined by Eq. (\ref{eq:problem_momentum}) satisfies the symmetry $\mathcal{I} \mathcal{H}(k)\mathcal{I}^{-1}=\mathcal{H}(-k)$, with $\mathcal{I}=\mathcal{P}_{y} \mathcal{K}$, where $\mathcal{P}_{y}$ is the parity operator $\mathcal{P}_{y} \psi_n=\psi_{-n}$, and $\mathcal{K}$ is the element wise complex conjugation operation. As a result of the symmetry protection, the Zak phase is quantized \cite{Benalcazar2017},  taking values of $\gamma_{n}=0$ or $\gamma_{n}=1$ depending on whether the $n$-th band is trivial or topological, respectively.

The Zak phase for the lower energy modes is indicated in Fig. \ref{fig:Bands}.  Here, $\gamma_{0}=1$ is provided for the lower energy band and $\gamma_{1}=0$ for the upper band,  and $\gamma_{2}=1$ for the next band. Consequently, we conclude that there are two topological band gaps in the proposed model. In contrast, under open boundary conditions, the edge-bulk correspondence implies that the non-trivial topology results in the presence of edge modes with energies within the topological gap. In the next section, we compare these theoretical predictions with micromagnetic numerical simulations, confirming the appearance of edge modes at the ends of the bc-DW.
Finally, it is worth noting that in the case $B_{y}=0$ and  $D/A\ll  w $, we have that $\epsilon \approx 0$, thereby the effective potential can be approximated, at first order in $\epsilon$, as $V_{\mathrm{eff}}(y)= - \epsilon \cos \left( \frac{D}{2 A}y+ \eta_{0} \right)$ (see Supplementary Material), which gives rise to a non-trivial topological structure with topological edge states, whenever the potential shift $\eta_{0} \ (\mathrm{mod}\ \pi)$ is nonzero \cite{Zheng2014}.
\begin{figure}[h!]
	\centering
	\includegraphics[width=\columnwidth]{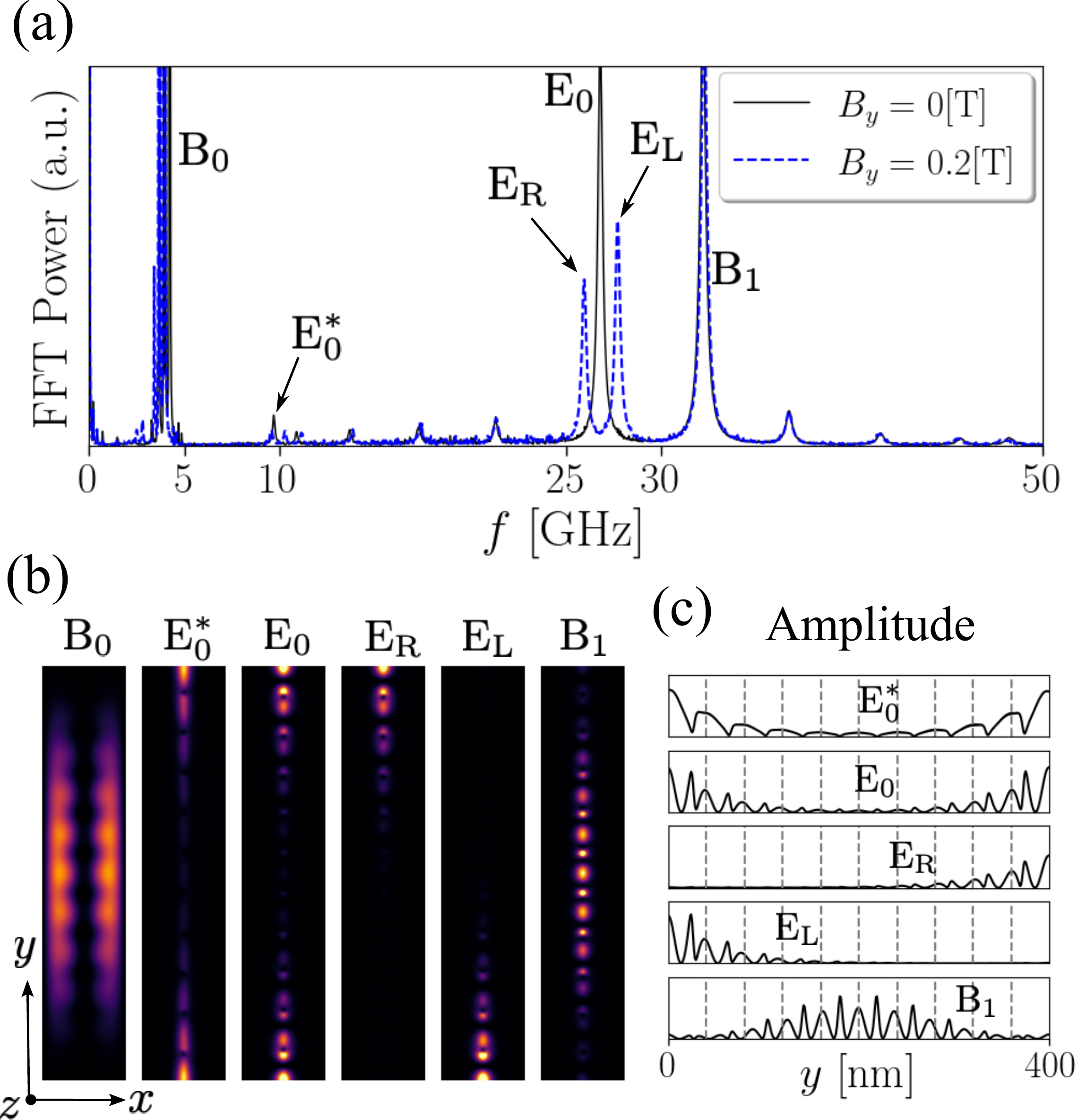}
	\caption{{Spin waves modes of a bc-DW,} with $N=10$ bimerons (see Fig. \ref{fig:Bimeron_Chain}). (a) Ferromagnetic spectrum for external magnetic fields  $B_{y}=0$ and $B_{y}=0.2$ T. For the case $B_{y}=0$, the observed modes are $\mathrm{E}_{0},\mathrm{E}_{0}^{*}, \mathrm{B}_{0}, \mathrm{B}_{1}$ , whereas for $B_{y}=0.2$ T, the modes are $\mathrm{E}_{L},\mathrm{E}_{R}, \mathrm{B}_{0}, \mathrm{B}_{1}$. (b) The local mode amplitude in color code. (c) The mode amplitude along the symmetry axis $x=0$ of the bc-DW.}
	\label{fig:Magnon_Modes}
\end{figure}

{\it Micromagnetic simulations.}-- The simulations were performed using the GPU-accelerated micromagnetic software package  $\mathrm{Mumax}^{3}$ \cite{Mumax3}, which solve Landau-Lifschitz-Gilbert (LLG) equation,
\begin{align}
\partial_t \bm{m}=-\gamma \mathbf{m} \times \bm{H}_{\mathrm{eff}}+\alpha \bm{m} \times \partial_t \bm{m}
\end{align}
where $\gamma$ is the electron gyromagnetic ratio, $\alpha$ is the Gilbert damping constant, and the effective field is given by $\mu_0 {\bm H}_{\text {eff }}=-\partial_{{\bm m}} \mathcal{E} / M_{\mathrm{s}}$. The dimensions of the geometry are $L_{x}=80$ nm and $L_{y}= 400$ nm. The cell size is $1 \times 1 \times 1$ \  nm$^{3}$. The magnetic parameters used in the simulations are as follows: exchange stiffness constant $A_{\mathrm{ex}}= 15$ pJ/m, saturation magnetization $M_{\mathrm{sat}}= 580$ kA/m, $D= 3$ mJ/m$^{2}$, easy-axis anisotropy along the $x$-direction from $K_{u}=500$ kJ/m$^{3}$, boundary anisotropy $K_{S}=1$ mJ/m$^{2}$, and the damping constant $\alpha=0.005$.  We initialize the magnetization of a bc-DW containing $N$ bimerons using the ansatz $\Theta(x,y) = \pi/2\ (1-\tanh(x/w))$ (domain wall far from the center) and $\Phi(x,y) = -2\pi N y /L_{y}$. The system is then relaxed to the equilibrium state. The energy of a bc-DW as a function of its number of bimerons is illustrated in the Supplementary Material.

The spin wave spectrum of a bc-DW is found by analyzing the power spectrum of the dynamical response of magnetization $\delta \boldsymbol{m}(\bm{r},t)= \boldsymbol{m}(\bm{r},t)-\boldsymbol{m}(\bm{r},t=0)$, under a uniform sinc magnetic field $\boldsymbol{B}(t)= B_{0}  \sin(2\pi f_{\mathrm{max}}(t-t_{0}))/(2\pi f_{\mathrm{max}}(t-t_{0})) \boldsymbol{y}$ of strength $B_{0}=1$ mT and the cutoff frequency of $f_{\mathrm{max}}= 50$ GHz, and $t_{0}=1$ ns. Then we determine the spatial FFT distribution as the Fourier image of each magnetic moment excitation $\delta \boldsymbol{m}_{\omega}(x,y) = \text{DFTt} (\delta \boldsymbol{m}(x,y, t))$, where DFTt is the discrete-time Fourier transform. Fig. \ref{fig:Magnon_Modes} shows the resonance frequencies and their corresponding spatial FFT amplitude distributions for the cases $B_{y}=0$ and $B_{y}=0.2$ T. The observed modes are categorized into two groups. (i) Bulk modes, displaying an extended amplitude on the geometry, here in both cases, $B_{y} = 0$ and $B_{y}= 0.2$ T, we see the modes $\mathrm{B}_{0},\mathrm{B}_{1}$; (ii) Edge localized modes, denoted by $\mathrm{E}_{0}^{*}, \mathrm{E}_{0}$, for the case $B_{y} = 0$, and $\mathrm{E}_{L}, \mathrm{E}_{R}$ if $B_{y}= 0.2$ T. We observe that the energy of the edge mode $\mathrm{E}_{0}^{*}$ lies within the observed band gap between 5 GHz and 10 GHz, while the energies of the edge modes $\mathrm{E}_{0}, \mathrm{E}_{L}, \mathrm{E}_{R}$ lie within the band gap between 25 GHz  and 30 GHz. These results support the predictions of our theoretical model, confirming the topological nature of the described modes. We can also note the appearance of energy splitting associated with the edge states $\mathrm{E}_{L}$ and $\mathrm{E}_{R}$ when $B_{y}$ does not vanish. We attribute this effect to the breaking of inversion symmetry, analogous to the energy splitting of edge modes in the Rice-Mele model \cite{Rice1982}. 

{\it Resilience of edge modes.}-- In this section, we explore the robustness of edge modes against the disorder effect. We introduce magnetic inhomogeneities into the system in the form of Voronoi grains of average size $20$ nm, such that each grain had a uniformly distributed anisotropy strength of $K_{u}(\bm{r})= K_{u}+ \chi(\bm{r}) \Delta K_{u}$, where $-1<\chi(\bm{r}) <1$ is a random number defined over each grain in the region. In our simulations, we set $K_{u}=500$ kJ/m$^{3}$ and $\Delta K_{u}=0.1\ K_{u}$.
\begin{figure}[h!]
	\centering
	\includegraphics[width=\columnwidth]{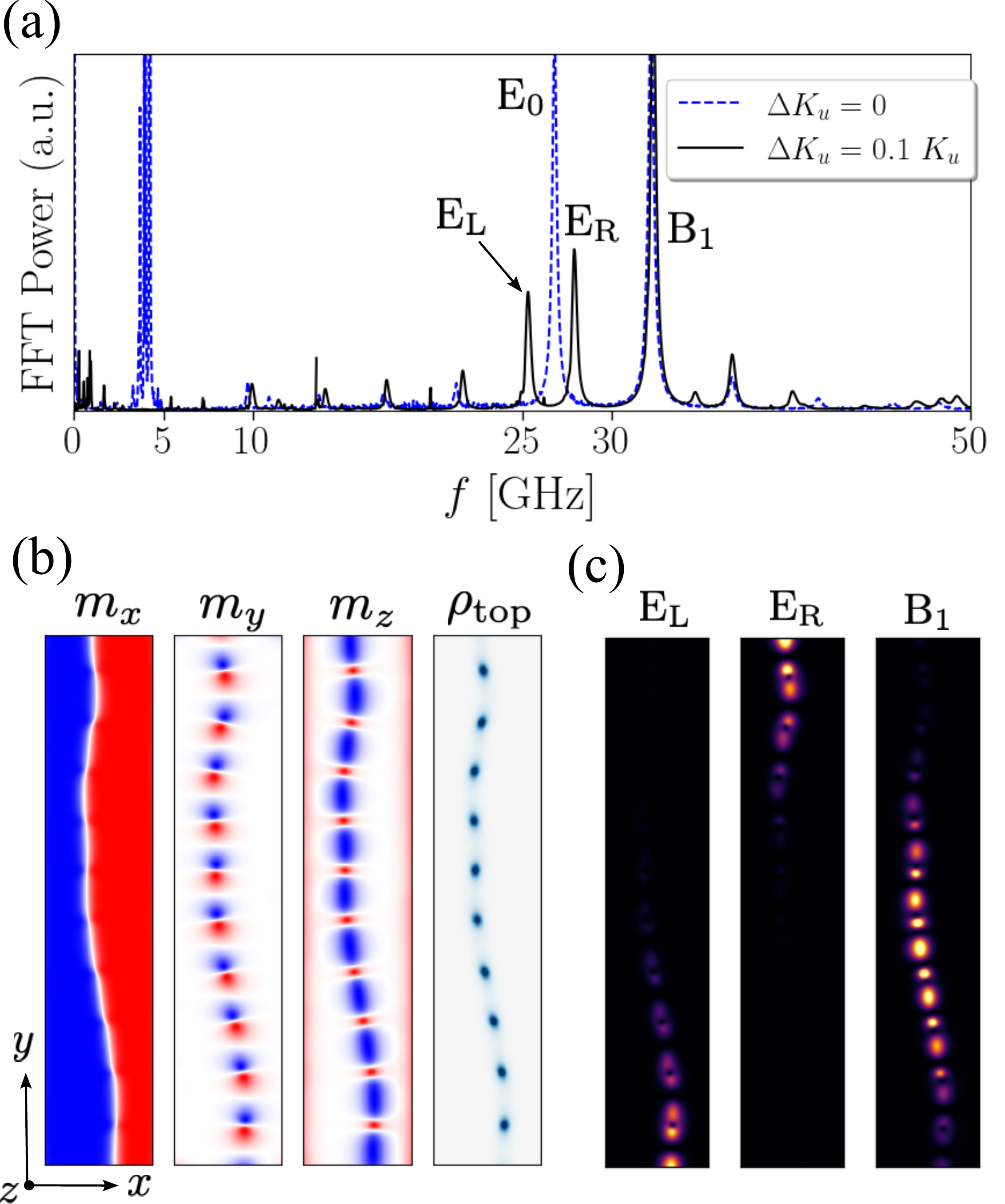}
	\caption{{Magnonics Modes in a bc-DW with disorder anisotropy}. (a) Ferromagnetic spectrum for the homogeneous case $\Delta K_{u}=0$ (it is the same as the case $B_{y}=0$ in the figure \ref{fig:Magnon_Modes}), and with anisotropy disorder $\Delta K_{u}=0.1\ K_{u}$. In the disordered case, we report the existence of the two edge modes $\mathrm{E}_{L},\mathrm{E}_{R}$. (b) Components $\bm{x},\bm{y},\bm{z}$ of the magnetization and the topological density charge for the disordered case. (c) The local mode amplitude in color code.}
	\label{fig: modes_disorder}
\end{figure}
The spectrum of the disordered system is shown in Fig. \ref{fig: modes_disorder}. Notably, the edge states remain stable even when the domain wall significantly deviates from the symmetry axis due to the random inhomogeneities.



{\it Conclusions.}-- We demonstrate that the domain walls of bimeron chains support topologically protected magnonic edge modes, making them candidates for magnon-based information transfer. A bc-DW serves as a remarkable example of a magnonic topological waveguide, which exhibits unique edge modes. As we show, these edge modes stand out for their remarkable resilience against a variety of external disturbances, making them particularly appealing for technological applications. This robustness is not an accident; it arises from the intricate design of the bc-DW, which creates a distinctive structure that supports these protected states.

The potential applications of these edge modes are vast, especially in the realm of nanotechnology. One of the most exciting possibilities lies in their use for the development of advanced nanodevices aimed at highly efficient data storage solutions built based on the localized magnon states with exceptional stability, which can effectively manage and transport information with increased reliability and efficiency. Thus, domain walls of bimeron chains represent an important advance in the search for innovative technologies in information storage and transmission.


{\it Acknowledgments.-} A.S.N. and R.E.T. acknowledges funding from Fondecyt Regular 1230515 and 1230747, respectively. C.S. thanks the financial support provided by the ANID National Doctoral Scholarship Nº21210450. M.A.C. acknowledges Proyecto ANID Fondecyt de Postdoctorado 3240112. E.S. acknowledges support from Dicyt-USACH 042331SD.




\bibliography{biblio}

\end{document}